%% file: coling_latex.tex
\pdfoutput=1

\documentclass[11pt]{article}

\usepackage[final]{coling}

\usepackage{times}
\usepackage{latexsym}
\usepackage{algorithm}
\usepackage{algorithmic}

\usepackage{newfloat}
\usepackage{natbib}
\usepackage{listings}
\usepackage{times}
\usepackage{latexsym}
\usepackage{hyperref}
\usepackage{booktabs}
\usepackage{multirow}
\usepackage{amsmath}
\usepackage{amsfonts}
\usepackage{amssymb}
\usepackage{color}
\usepackage{tikz}
\usepackage[edges]{forest}
\definecolor{hidden-draw}{RGB}{20,68,106}
\definecolor{hidden-pink}{RGB}{255,245,247}
\usepackage{xcolor}  
\usepackage{newunicodechar}
\usepackage{bibentry}

\usepackage{graphicx}
\usepackage{textcomp}
\newunicodechar{￥}{\textyen}
\usepackage[T1]{fontenc}

\usepackage[utf8]{inputenc}

\usepackage{microtype}

\usepackage{inconsolata}
\usepackage{graphicx}

\title{A Framework for Effective Invocation Methods of Various LLM Services}

\author{Can Wang,\, Dianbo Sui,\, Bolin Zhang,\, Xiaoyu Liu,\, Jiabao Kang,\, Zhidong Qiao,\, Zhiying Tu\thanks{*Corresponding Author.} \\
  Harbin Institute of Technology (HIT), Weihai, Shandong, China \\
  \{23B903072\}@stu.hit.edu.cn \\
  \{suidianbo, brolin\}@hit.edu.cn \\
  \{24S130270, 23B903065, 23S136230\}@stu.hit.edu.cn \\
  \{tzy\_hit\}@hit.edu.cn \\
}

\begin{document}

\maketitle

\begin{abstract}
Large Language Models (LLMs) have shown impressive abilities in solving various natural language processing tasks and are now widely offered as services. 
LLM services enable users to accomplish tasks without requiring specialized knowledge, simply by paying service providers.
However, numerous providers offer various LLM services with variations in pricing, latency, and performance. 
These factors are also affected by different invocation methods, such as the choice of context and the use of cache, which lead to unpredictable and uncontrollable service cost and quality.
Consequently, 
utilizing various LLM services invocation methods to
construct an effective (cost-saving, low-latency and high-performance) invocation strategy that best meets task demands becomes a pressing challenge. 
This paper provides a comprehensive overview of methods help LLM services to be invoked efficiently.
Technically, we define the problem of constructing an effective LLM services invocation strategy, and based on this, propose a unified LLM service invocation framework.
The framework classifies existing methods into four categories: input abstraction, semantic cache, solution design, and output enhancement, which can be used separately or jointly during the invocation life cycle.
We discuss the methods in each category and compare them to provide valuable guidance for researchers.
Finally, we emphasize the open challenges in this domain and shed light on future research.
\end{abstract}

\section{Introduction}

Large Language Models (LLM) are becoming a fundamental tool for various natural language processing tasks~\cite{28}, as they have shown amazing emergent abilities, like in-context learning~\cite{30}, multi-step reasoning~\cite{FuPSCK23}, instruction following~\cite{LouY24} and tool learning~\cite{huang2024practical}. Due to commercial reasons, the potential risk of misuse and expensive tuning cost, LLMs, such as GPT-3~\cite{12}, GPT-4~\cite{gpt4} and Claude\footnote{\url{https://claude.ai/}}, are usually released as LLM services through application programming interface (API) instead of open sourcing model weights, which is called Language Models as a Service (LMaaS)~\cite{lmturk}. 
By accessing these powerful LLMs as services through their opened API, novice users do not need to possess
extensive computational resources and expertise in deep learning, as they can solve the tasks of interest
by crafting task-specific input queries. 

\begin{figure}[t]
  \centering
  \includegraphics[width=0.9\linewidth]{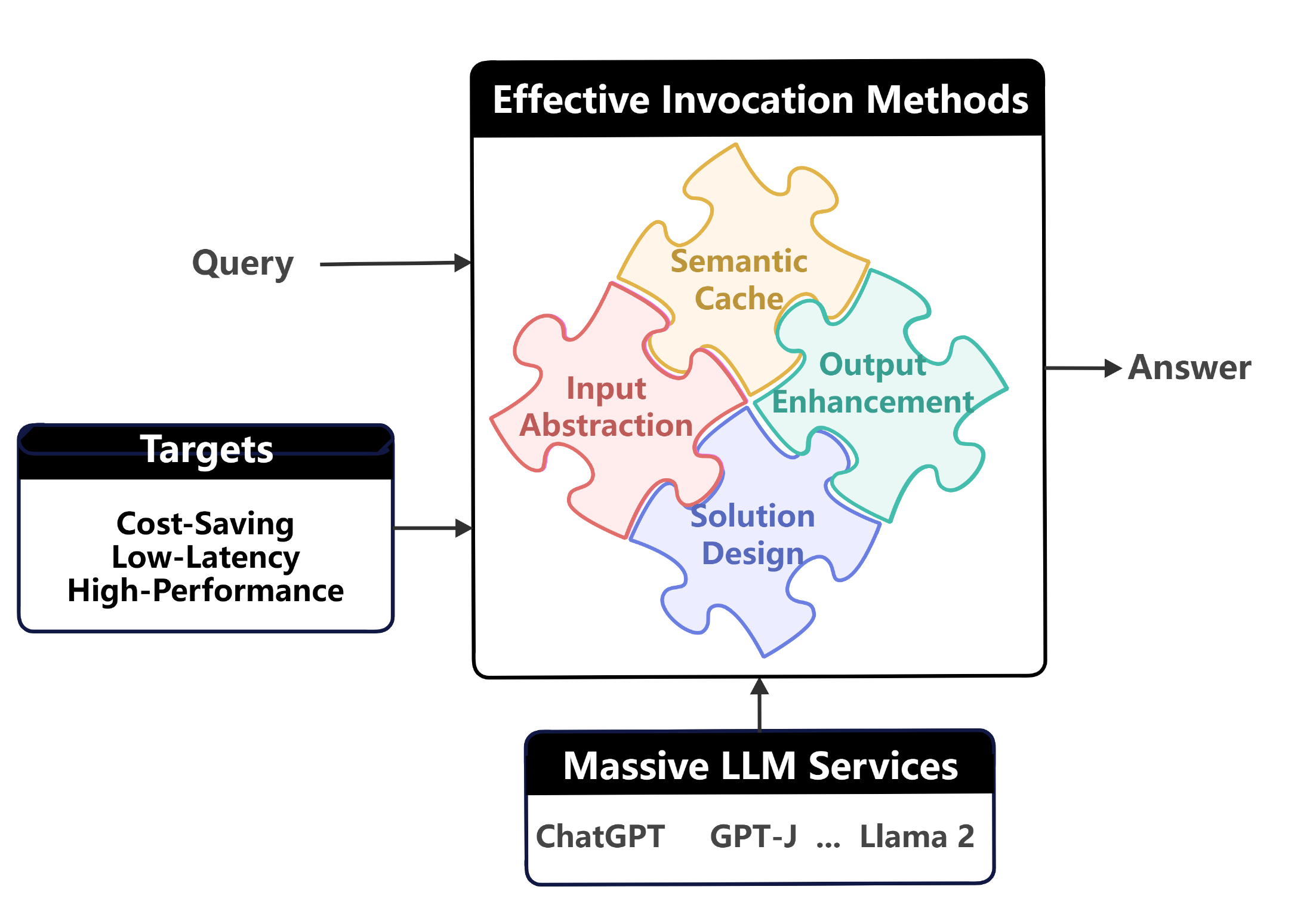}
  \caption{Vision of effective invocation strategy construction for various LLM services.}
  \label{Vision}
\end{figure}

However, invoking LLM services is not free and using them for high-throughput applications can be very expensive. Estimated by
Claudia Slowik\footnote{\url{https://neoteric.eu/blog}}, a business supporting 15,000 customer interactions with text-davinci-003 could have a monthly cost exceeding \$14,400. 
We present the cost of using 25 different LLM services from some top-tier providers (as shown in the Appendix) and find that the costs of different LLM services can vary by up to two orders of magnitude. For instance, the input cost for 1 million tokens is \$10 for OpenAI’s GPT-4 but only \$0.2 for Mistral 7B hosted by Textsynth.

In addition to cost considerations, various factors, including response time and performance for the same input query, can also impact the user experience using LLM services.~\citet{27} and~\citet{29} find that different languages, prompt methods or the inclusion of simple enhancements can also lead to notable alterations in performance. Meanwhile,~\citet{1} discover that affordable LLMs often complement expensive ones. For instance, on the CoQA~\cite{CoQA} dataset, GPT-4 makes errors in approximately 11\% of the questions, while the more affordable GPT-J provides the correct answers.

From the fact that the heterogeneity in pricing does not necessarily correlate with the user experience, it is a great need to explore effective invocation methods for LLM services in practice. As shown in Figure~\ref{Vision}, 
we expect to make use of various LLM services to construct an effective invocation strategy, which can be easily adjusted and reused according to the users' different invocation targets.
To this end, we provide a comprehensive study of the development and recent advances
on effective invocation methods of LMaaS. In detail, we first formalize the task of constructing an effective invocation strategy as 
a multi-objective optimization problem~\cite{gunantara2018review}
, which entails simultaneous consideration of latency, performance, and cost factors. Then, we propose a taxonomy to provide a unified view on effective invocation methods of LMaaS where existing methods are categorized into: input abstraction, semantic cache, solution design, and output enhancement. These four categories can be flexibly combined and unified into a flexible framework. Finally, we highlight the challenges and potential directions.

The contributions of this survey can be concluded as follows:
\begin{itemize}
    \item \textbf{Comprehensive Taxonomy}. 
    We define the problem of constructing an LLM services effective invocation strategy mathematically. Based on this, we propose the taxonomy in Figure~\ref{taxo_of_icl}, which categorizes existing methods from four different aspects: input abstraction, semantic cache, solution design, and output enhancement.
    \item \textbf{Flexible Framework}. As shown in Figure~\ref{fig:framework}, we propose a framework that unifies the four categories of methods. Our framework connects different categories during the LLM service invocation life cycle, allowing each of them to be used separately or jointly.
    \item \textbf{Related Resources}. To facilitate the methods of this task, the price rules of popular LMaaS products are shown in Table~\ref{price}. The anonymous GitHub repository presents the existing methods available, with a demo website implementing our framework.\footnote{\url{https://anonymous.4open.science/r/Effective-strategy-for-LMaas-BF83}}  
\end{itemize}


\section{Background}
\label{Background}
In this section, we first formalize the problem of constructing an effective invocation strategy of various LLM services.
Then, we explain the problem definition from the perspective of the LLM service invocation lifecycle, which divides the invocation process into three phases. According to the different use phases and purposes when construction, we propose our taxonomy and framework. 

\subsection{Problem Definition}

\input{taxonomy.tex}

In our topic, the problem is defined as how to construct an effective (cost-saving, low-latency and high-performance) invocation strategy $s$ given a task $T$ among various LLM services. The given task $T$ consists of multiple identical query-answer pairs, represented as $T=\{(q_1,a_1),(q_2,a_2),... ,(q_n,a_n)\}$, where $q_i$ represents input query and $a_i$ represents output answer.
First, we consider a fixed LLM service $M$ published through API. Input a query $q$, the process of obtaining the response $\tilde{a}$ by invocation of the LLM service $M$ can be represented as: 

\begin{equation}
\label{eq1}
\tilde{a} = M(q).
\end{equation}


To characterize the concerned factors for the construction of effective invocation strategy with a given query $q$ and LLM service $M$, we use three metric functions: latency $f_{l}(M, q)$, performance $f_{p}(M,q)$, and cost $f_{c}(M,q)$. These three functions are fixed values in a specific practical invocation and can be estimated using certain methods. 
For example, $f_l$ could be a function of the length of the input and output sequences. $f_p$ often uses a metric function $r(\cdot,\cdot)$ to compare the difference between $a$ and $\tilde{a}$. While $f_c$ involves two different pricing parts, input cost and output cost, 
we adopt the definition of the prompt length multiplied by the token price as
shown in the Eq.~\ref{eq2}, where $\alpha_i$ is a constant representing the unit price.


\begin{equation}
\label{eq2}
f_c \triangleq \alpha_1||q|| + \alpha_2||\tilde{a}|| +\alpha_3
\end{equation}



Based on that, we extend a single LLM service to $K$ different LLM services $M_s = \{M_1,M_2,...M_k\}$. Our problem is formalized as in Eq.~\ref{eq3}, where in the search space $S$, we seek an optimal invocation strategy $s$ that minimizes latency $f_{l}$, maximizes performance $f_{p}$, and minimizes cost $f_{c}$ on task $T$. The optimal strategy $s$ includes a sequence of selected LLM services, represented as $s = \{M_i\}, i\leq K$, offering flexibility in choosing one or multiple services in a specific order.

{\small
\begin{equation}
\label{eq3}
\begin{aligned}
\text{Minimize } & \mathbf{F}(s) = \begin{bmatrix} f_l(M_i,q_j)\\ -f_p(M_i,q_j)\\ f_c(M_i,q_j) \end{bmatrix}\\
\text{subject to } & M_i \in s , q_j \in T 
\end{aligned}
\end{equation}
}


This is a multi-objective optimization problem, and we solve it using simply weighted sum or other methods.
In the construction strategy of a specific invocation, 
constraints may be introduced. 
For example, in the scenario of limited funds, the cost $f_c \leq C$ is used as a condition to obtain a strategy with high-performance $f_p$ and low-latency $f_l$, where $C$ is the maximum cost that can be used.

\subsection{LLM Services Invocation  Taxonomy and Framework}




Following the idea that building an effective invocation strategy requires an understanding of the key resources involved in the LLM service life cycle~\cite{16}, we organize our framework by dividing the LLM service invocation into three phases: before invocation, during invocation, and after invocation. 
According to the different phases and purpose, the taxonomy divides the methods to effectively invoke LLM services into four major categories, which are connected in a sequential way in our proposed framework to optimize the common goal in Eq.~\ref{eq3}.

\begin{figure*}[t]
  \centering
  \includegraphics[width=0.98\linewidth]{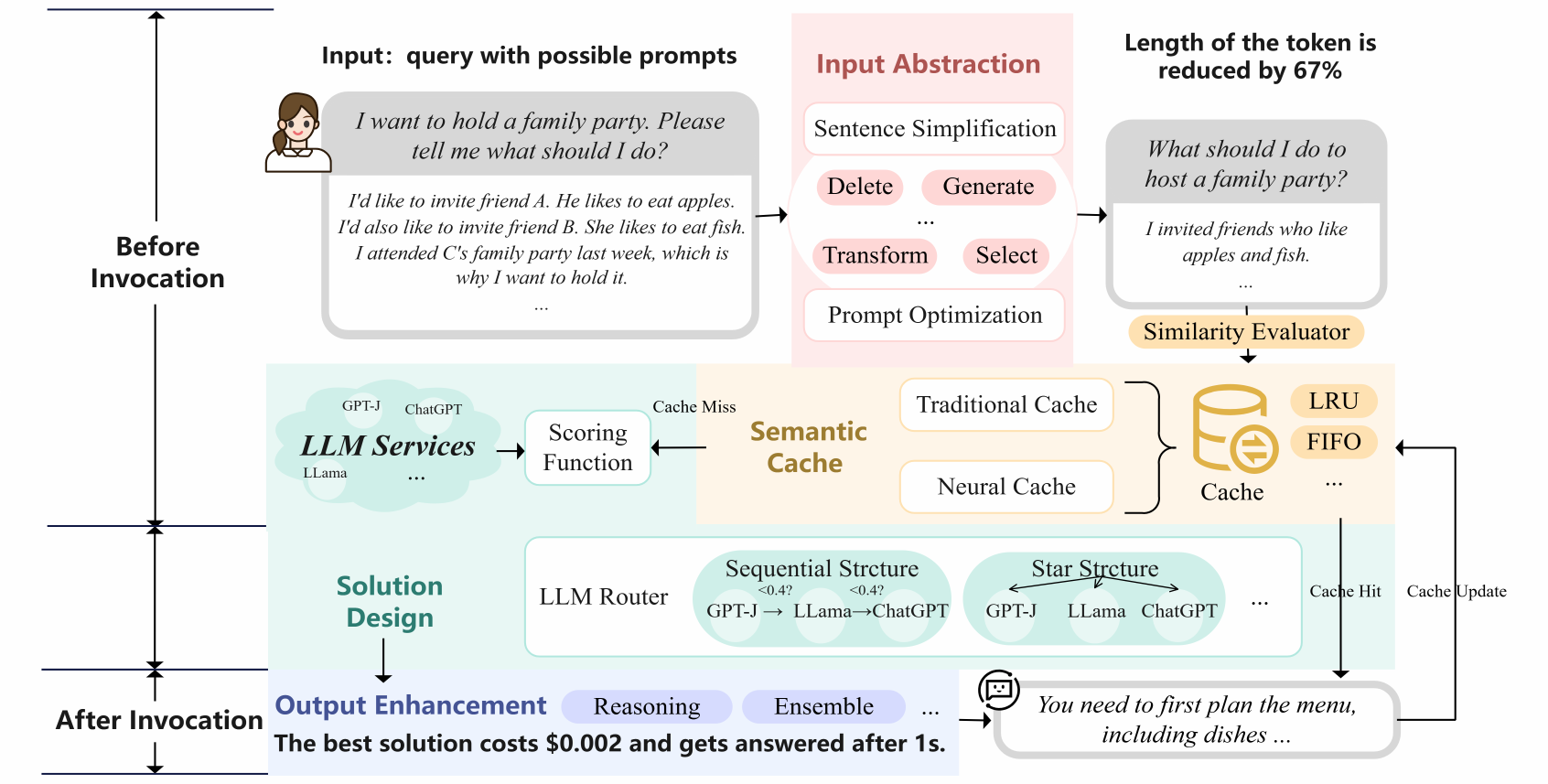}
  \caption{LLM services invocation framework, shown by the phase of invocation.}
  \label{fig:framework}
\end{figure*}

Before invocation, 
processing the input query $q$ that a user entered to express more meaningful information in a more concise way is the first step to construct the effective invocation strategy. The related methods are summarized as \textbf{input abstraction} (Section~\ref{section:abs}), which can be further divided into \textbf{sentence simplification} and \textbf{prompt optimization} according to the different ways.
\textbf{Semantic cache} (Section~\ref{section:Cac}) is also an important method to improve service performance, reduce latency and cost before invocation, which is divided into \textbf{traditional cache} and \textbf{neural cache} according to different structures. The semantic cache checks whether there is a semantically similar query $q$ in cache. If so, it directly returns the previous answer $\tilde{a}$, and otherwise it goes into the invocation phase.

During invocation, \textbf{solution design} (Section~\ref{section:int}) aims to construct the best invocation solution $s$ by leveraging the complementary capabilities of various LLM services $M_i$. It evaluates LLM services $M_i$ with a given query $q$, and the method of evaluation is called \textbf{scoring function}. It can also be done before the invocation. For example, the estimation of $f_c$ can be used to guide the design of a cost-saving solution. During the invocation phase, the \textbf{LLM router} is used to organize routing structures between LLM services according to the estimated results. Through different routing structures, the advantages of different services are utilized to build optimal strategies for users.

After invocation, \textbf{output enhancement} (Section~\ref{section:Out}) focuses on the information returned to the user. The output $\tilde{a}$ is enhanced to improve clarity and accuracy, allowing it to better meet different targets and enhance the overall user experience. Output enhancement can be implemented in two main ways: \textbf{thought reasoning} and \textbf{ensemble learning technology}. 
In addition, the input and output of this invocation are stored into the semantic cache for future invocations.

In summary, we categorize effective invocation methods for LLM services into four groups, as shown in Figure~\ref{taxo_of_icl}, based on different phases of use and their construction purposes.
We discuss these categories and compare the methods in later sections, providing advantages, disadvantages, and applicable scenarios. We propose the LLM services invocation framework that can unify these methods as illustrated in Figure~\ref{fig:framework}, where different categories of methods can be used separately or jointly.

\section{Input Abstraction}
\label{section:abs}

Input abstraction is designed for better performance at lower cost and latency to invoke a given LLM.
The generalization and in-context capabilities allow LLM services to obtain good answers on untrained samples~\cite{30}, and input content directly affects the cost, latency or performance of the service. 
For example, concatenating the prompt ``Just tell me the option" with the question as input to the LLM will generate shorter output, reducing invocation cost and latency. However, it may cause the LLM to lose its ability to think step by step, resulting in performance degradation.

We group the methods into two categories based on different goals, sentence simplification and prompt optimization. 
The input query $q$ typically consists of a question (representing the user's task) and multiple prompts (optional information to aid in task completion).
Sentence simplification reduce the length of the input query without changing the semantics, while prompt optimization ensures the quality of the query and improves the performance of the invocation by optimizing prompts.

\subsection{Sentence Simplification}
\label{section:Sen}

Sentence simplification is the process of making input more concise and simple while retaining its core meaning by modifying, removing, or replacing words, phrases, or structures in a sentence.
The process is similar to the summarization task, and many methods used in summarization can be applied~\cite{31,32,33,34}. We collate the methods available for LMaaS, and classify them into extractive and generative methods based on whether derived entirely from the original input.


\textbf{Extractive methods.} Extractive methods select sentences from long original input by extracting key sentences or phrases, where the content is entirely sourced from the origin. Pruning semantically irrelevant tokens according to the relevance to the contexts is an effecitve approach~\cite{38}. Iterative deletion and substitution of tokens are another methods~\cite{8,2}, removing unimportant tokens based on the attention mechanism.

Extractive methods are straightforward and efficient, making it convenient for the real-time use. However, the extractive methods may ignore some global information. 
And it has limitations in some tasks such as language translation, as it cannot discern which parts need to be translated or deleted.

\textbf{Generative methods}. The generative methods refer to rewriting based on the original input, allowing for the generation of new words. Language encoding is a simple processing method. ~\citet{27} conduct extensive experiments with different languages and tokenizers, where the cost varied by up to 5 times. AE.studio\footnote{Prompt Reducer: \url{https://www.promptreducer.com/}} employs encryption to provide an online platform that sacrifices readability, reducing the length of input tokens by half. Besides, utilizing fast and low cost generative models~\cite{52} presents a viable option for sentence simplification.

Generative methods are flexible, as the generated sentences contain less redundant information while preserving the main content. However, it may introduce grammatical or factual errors. 

\subsection{Prompt Optimization}
\label{section:Pro}

Prompt optimization is the adjustment of user-provided input prompts to guide LLM to produce more accurate, useful, and tailored output. The effectiveness of prompt optimization stems from the LLM's ability to learn from few-shot demonstrations~\cite{40}, where appropriate prompts can complement the context of a task, highlight key information, or improve the explainability.

Based on the different granularity of optimization objective, we distinguish two types of prompt optimization methods. Prompt selection selects or combines some sentences in prompts to guide invocations more effectively.
Prompt augmentation is concerned with the quality of the content and aims to maximize the potential of the context.



\textbf{Prompt selection.} Prompt selection selects the most meaningful prompt from possible prompts to accurately guide LLM. It removes the interference of irrelevant prompts, and helps in efficient invocation.~\citet{37} select representative samples that can be highly beneficial in few-shot tasks.
~\citet{41} combine various methods involving instructions, examples, and additional context to propose a more compact method for providing prompts in dialogues.~\citet{4} consider the concatenation of prompts and retrieves the most important $k$ sentences, enabling the shared use of prompts for similar questions.

Prompt selection directly guide LLM to focus on specific information and understand user needs more accurately without too much personalization. 
However, this approach cannot maximize the potential of LLM capability for complex prompts because no additional knowledge is introduced.

\textbf{Prompt augmentation.} Prompt augmentation considers the understanding ability of LLM to elicit more accurate and desirable responses. 
Knowledge retrieval is a direct method of augmentation, as it helps achieving a comprehensive understanding during the invocation.~\citet{39} respond the limitations of factual knowledge in LLM and optimizes the reasoning process with minimal retrieval cost.~\citet{11} and~\citet{37} propose black-box fine-tuning methods to optimize continuous prompts using non-derivative methods. Model alignment~\cite{43} and chain of thought reasoning~\cite{44}, are also key focuses in prompt optimization.

The improvement in invocation performance through prompt augmentation is significant, despite it may resulting in more complex processing procedures. However, the general method is difficult to explore, requiring professional knowledge.

\section{Semantic Cache}
\label{section:Cac}
Semantic cache is an approach to improve LLM invocation efficiency and performance by storing and quickly retrieving information. Unlike traditional data cache, semantic cache focuses on storing high-level semantics information such as meaning, relationship, rather than just raw data. The semantic cache is checked before the LLM service is invoked. If cache hit, the output given by the cache is returned, avoiding subsequent costly invocations while responding faster.
With the gradual increase in the scale of LLM, the semantic cache plays a more important role in accelerating computation, reducing data transmission costs, and supporting high concurrent requests~\cite{26}, providing users with low-latency, high-performance, and cost-saving services.

There are two typical structures for implementing semantic cache in LMaaS, and unlike other subsections, these two structures generally cannot be used jointly. Traditional caches use key-value pairs for storage and retrieval, returning the same value for similar input. Neural caches, on the other hand, use neural networks to respond in a predictive manner, learning semantic relationships between inputs without relying on a fixed storage structure.

\subsection{Traditional Cache}
\label{section:Tra}
The current paradigm of traditional cache consists of three parts~\cite{5}: the cache manager, similarity evaluator and post processor. The cache manager is responsible for storing content in the form key-value pairs, and managing eviction. The similarity evaluator is used to determine if any of the keys in the cache match the input query. The post processor organizes the final response to be returned to the user. If no similar query is found in the cache, the LLM service is invoked by the post processor to generate the output and then the generated output is stored in the cache.

\citet{5} represents a typical implementation of traditional cache, which utilizes question embedding for similarity matching and provides various matching methods. The open-source application Zep\footnote{\url{https://github.com/getzep/zep}} supports storage LLM applications, storing information in the database. Through theoretical proof,~\citet{47} introduce the cache scheme with minimum expected cost considering the query frequency. Besides, methods for query and conversations cache~\cite{13,46} can be migrated to LLM services.

Implementing traditional cache is usually simple, requiring only basic data structures such as hash set. This approach is general, but it may not capture semantic similarity between inputs because it relies heavily on the key matching.

\subsection{Neural Cache}
\label{section:Neu}
Neural cache uses neural networks or deep learning models to learn and store data representations. Neural cache maps input data into a high-dimensional space by learning the representation of the data, which can capture the semantic similarity of the input. Unlike the compositional paradigm of traditional cache, neural cache has no specific structure.

\citet{6} train a student model using T5-base\footnote{\url{https://huggingface.co/docs/transformers/model_doc/t5}} for providing early feedback in classification tasks. To address the semantic cache missing issue,~\citet{36} generate similar input to hit the cache as much as possible. Furthermore, a retrieval-based dialogue response selection model~\cite{13} also can serve as an alternative choice for neural cache. 

The neural cache often outperforms the traditional cache, especially in domain-specific problems. However, its implementation and updates can be relatively complex. As such, it is important to carefully consider the effectiveness of the cache to avoid incurring unnecessary resource waste.

\section{Solution Design}
\label{section:int}

Solution design 
considers different scenarios and targets, dynamically selecting one or more LLM services that are most suitable for the invocation, and organizing them to provide flexible and efficient solutions. The solution design helps to select LLM services that best solve the task, taking advantage of LLM services' heterogeneous costs and performance. When new queries arrive or requirements change, the solution can be flexibly updated to achieve optimal performance and cost savings.

Solution design has two main parts working together to achieve dynamic LLM service selection and routing. The scoring function is responsible for evaluating the performance of each available LLM service, which reflects the concerned factors for invocation such as quality and speed. The router, based on the evaluation results of the scoring function, performs query routing between services, and selects the appropriate one in a dynamic manner.

\subsection{Scoring Function}
\label{section:Eva}
The scoring function is a comprehensive evaluation of LLM services given a specific task, considering both targets and scenarios, which is used to guide the routing path in the solution. The scoring function may be influenced by multiple factors, such as response time, query cost, and accuracy of answers. The scoring function plays a decision-making role, and helps to understand the relative performance of each LLM service. The scoring function can be implemented in two different ways.

\textbf{Defined metrics.} Defined metrics provide a measurable way to achieve direct quantification of factors that affect invocation. For instance, accuracy in classification tasks, BLEU score in generation tasks, packet loss in web, and quality of service (QoS) are all applicable indicators.~\citet{6} use interval sampling and predictive entropy to determine whether to invoke LLM services. Considering three sources of consistency, decision-making for LLM services is performed through sampling and voting~\cite{10}. Calculating the cost expectations between two models,~\citet{47} extend the decision when invoking multiple LLM services. Reward ranking from answers provided by different services is used as an evaluation criterion by~\citet{51}, incurring minimal computational cost in the solution.

The defined metrics are intuitive and easily understandable, which are often based on statistical data or experiments, being less susceptible to subjective factors. However, setting thresholds for evaluating the quality of LLM services can be challenging and may not adapt well to the dynamic and changing environment. Additionally, certain complex factors may be difficult to capture with defined metrics, leading to limitations in scoring.

\textbf{Scorers.} A scorer is a tool for scoring LLM services based on metrics that cannot be defined by a particular formula. The scorer utilizes prior knowledge, training data, or rules to provide scores in a less interpretable manner, typically using smaller neural networks~\cite{1}. For example, ALBERT~\cite{lan2020albert} is used as a scorer~\cite{7}, with the query and predicted output as $x$, and the accuracy of invocation as $y$. Using DistilBERT~\cite{sanh2020distilbert} as a scorer, with query and model ID as $x$,~\citet{53} predict whether an LLM can solve the problem. A comparison of LLM performances on different benchmark datasets is conducted, with~\citet{48} modeling it as a binary selection problem and providing guiding suggestions. 
For specific tasks, such as the predictor of execution results in the code generation task~\cite{48}, the classifier of query difficulty in the task of question and answer~\cite{15,50}, and estimator of LLM service capability in the dataset benchmark test task~\cite{53} are all reasonable scorers.

Compared with metrics defined by formulas, the scorers can be updated based on real-time data and feedback, demonstrating strong generalization across different scenarios. However, it is equivalent to using a more powerful model and incurring its training and usage costs. Moreover, it still requires labeled examples, which is applicable only when the query dataset is larger than the training dataset.

\begin{figure*}[!t]
  \centering
  \includegraphics[width=0.9\linewidth]{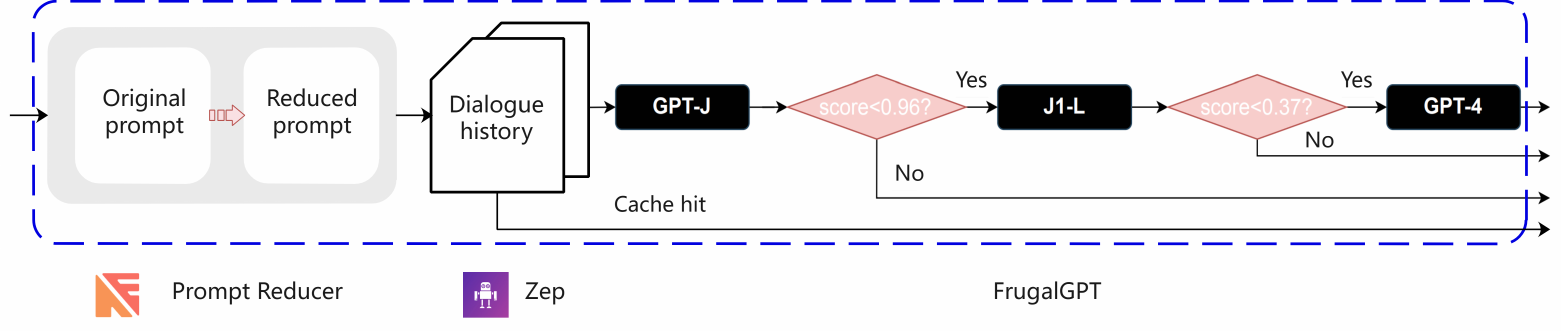}
  \caption{A simple invocation strategy composed of existing methods, using Prompt Reducer in input abstraction, Zep in semantic cache, FrugalGPT in solution design, and nothing in output enhancement.}
  \label{fig:case}
\end{figure*}

\subsection{LLM Router}
\label{section:Rou}
The LLM router emphasizes the organizational structure between LLM services, connecting multiple independent services in a specific order logically. It focuses on constructing a flexible and reusable solution to address continuously changing queries or targets. Depending on the scoring function and position used, the LLM router can construct various target-oriented solutions, such as cost-oriented or performance-oriented solutions.

\textbf{Sequential Structure}. Sequential structure is the simplest structure, which selects one or several available services from the extensive pool of LLM services and invokes them in sequence. 
The scoring function is used to decide whether to accept the answer or proceed to the next step~\cite{1}. The number of models is typically limited to three in a sequential structure, and possible options are determined through permutation, with pruning techniques applied~\cite{6}. The use of small models as a cache, with large models being invoked in sequence when cache misses occur, can also be regraded as a fixed sequential structure~\cite{10}. For problems like code generation~\cite{48}, an initial response is obtained using a cost-saving LLM, which is tracked as context for subsequent queries.

This structure is simple and effective, and a limited number of permutations can be searched quickly in the entire space. However, the sequential structure may result in the invocation of all LLM services in the sequential structure. Thus, it is difficult to extend, because all LLM services need to be rearranged when adapting to new requirements.

\textbf{Other Structure.}
Parallel structure, similar to the bagging in machine learning, can enhance the performance of LLM services, with task decomposition and merging being key aspects~\cite{49}. Star structure, as seen in~\citet{7,51}, involves decision-making by a meta-model, allocating the current query to the most suitable model. For the third category of unsolvable queries, pruning is applied by~\citet{50} to prevent unnecessary costs for particularly challenging problems. The tree structure is considered to be promising, combining advantages of both star-shaped and sequential structures. It initially routes query to the most probable branch, and then invokes services in sequence. Additionally, certain selection solutions specific to HTTP services~\cite{55,56} are also noteworthy to be explored.


Other structure offers more flexible and extensible solutions, while also introducing complexity into the routing process. 
Moreover, designing and implementing these structures can be challenging, particularly when dealing with dynamic query environments or evolving system requirements
\section{Output Enhancement}
\label{section:Out}
Output enhancement refers to the process of further optimizing and adjusting the output generated by the invocation. This process improve the syntactic correctness, semantic accuracy, and overall fluency of the generated output to meet needs of the user under the specific scenario. Depending on the method, output enhancement can be classified into thought reasoning and ensemble learning.

The thought reasoning focuses on the internal reasoning process of the LLM to improve the organization and explanation of outputs. Ensemble learning technology emphasizes the cooperation of multiple service invocations. By combining the strengths of different models, it enhances the stability and robustness of the LLM, and reduces the bias and error of the output of a single service.

\subsection{Thought Reasoning}
\label{section:Tho}
Thought reasoning improve reasoning clarity and logical flow in outputs by changing the path of reasoning inside LLM services. ~\citet{40} guide an LLM to give concise answers using carefully designed prompt, reducing unnecessary output tokens. ~\citet{60} use X-of-Thought to refer to the chain of thought in a broad sense, systematically organize the current research according to the classification of methods. In addition, work on model alignment identifies~\cite{58} and reduces the bias or unsafe behavior that LLM may produce in inference, reducing the need for subsequent human intervention.

The thought reasoning is easy to implement and obviously enhances the logicality and organization of the output through simple prompt optimization. However, thought reasoning relies on carefully designed prompts, and the effect is limited by the ability of the underlying LLM service.

\subsection{Ensemble Learning}
\label{section:Ens}
Ensemble learning is a powerful technique that combines multiple LLM services to improve performance over individual LLM services.~\citet{64} and ~\citet{65} discusse the work and challenges of model merging, combining structure or knowledge from multiple models into a single model.
To solve domain-specific tasks, ~\citet{63} enhance medical question answering capabilities based on boosting and clustering methods, and~\cite{9} aggregate responses from multiple low-cost LLM services to improve the precision of multi-label prediction tasks.

By combining the predictions of multiple LLM services, ensemble learning techniques significantly improve the accuracy and stability of the output, reduce the risk of overfitting, and flexibly adapt to multi-domain tasks. However, this approach requires more computational resources and time, leading to high costs, while the complexity of the model also makes the system difficult to debug and interpret.



\section{Conclusion and Challenges}
\label{Frontiers}

In conclusion, this paper provide a comprehensive overview of effective invocation methods in the realm of LMaaS. 
Through the establishment of a taxonomy, we categorize existing methods into four categories: input abstraction, semantic cache, solution design, and output enhancement. 
We formalize the problem of effective LLM services strategy construction, and propose an LLM services invocation framework. Methods of different categories in the framework can be used separately or jointly to form the effective strategy invocation that are low-latency, high-performance, and cost-saving.


Most existing methods focus only on one category in the framework, and the methods in different categories can be used as plugins. A practical example of a simple invocation strategy built from three existing methods to save money is shown in Figure~\ref{fig:case}. In addition, other factors may also be important when calling LLM services, including but not limited to interpretability, security, automation rate, etc. Our framework is open and flexible, allowing for easy expansion to these aspects.
We look forward to future research further advancing the field, and here are some open challenges.

\textbf{Input Abstraction.}
One of the main challenges faced in the input abstraction category is the processing of multi-modal input~\cite{100}. More comprehensive and balanced methods~\cite{125} are needed to optimize multiple types of input such as text, image, and speech. Input abstraction methods for dynamically changing queries are also worth exploring, such as real-time data streaming~\cite{101} or user interaction with the system. 

\textbf{Semantic Cache.}
In the semantic cache category, how to design and select cache methods~\cite{102} more efficiently to accommodate different inputs and queries is the main challenge faced in traditional cache, while semantic representation~\cite{104} can be achieved by neural cache. The study of more efficient algorithms for cache storage and update~\cite{126} is also a technique worth discussing.

\textbf{Solution Design.}
In terms of solution design, a quantitative evaluation of LLM services~\cite{103} is an extension of the scoring function, which adaptation and interpretability need to be paid more attention to in the future. The LLM router needs to focus on designing more powerful service integration methods that not only focus on a task, but also take into account requirements of different resources~\cite{109}. 

\textbf{Output Enhancement.}
The importance of output enhancement is gradually seen by people. The balance between specification and diversity~\cite{127} of output is a key issue. When a task is completed, the user's satisfaction is an important indicator to measure the quality of service, and future research may focus on building more intelligent and user-oriented~\cite{120} output enhancement methods.

\textbf{Other Challenges.}
Basic work such as qualitative description and quantitative comparison in experiments still has a gap to be filled, and the lack of baselines results in no uniform standard for the comparison of the LLM services invocation methods. Futher studies, such as how to choose the tokenizer~\cite{105} with the shortest input, how to set the best suitable size of cache~\cite{106}, and the choice of different pricing methods for the LLM service, need to be explored. In additionally, We specifically call for attention to fairness~\cite{107} and privacy issues~\cite{108,110} of LMaaS. 

\section*{Limitations}
First, in defining the construction of an effective invocation strategy for LLM services, we select three key factors and model the topic as a multi-objective optimization problem.
It is not comprehensive because there are many other factors in the actual invocation of LLM services, which also affect each other.
Second, we summarize the effective LLM service invocation methods, aiming to provide a general framework to help users build the cost-saving, low-latency and high-performance invocation strategy. However, some of these methods are limited because we do not consider whether they can be directly applied to black-box LLM services that are only published through APIs. 
Furthermore, due to the variety of experimental datasets and evaluation metrics, we are unable to conclude a unified baseline for this topic, which would be a strong support for future research.


\section*{Acknowledgements}
The work is supported by 
the National Key R\&D Program of China (Grant No.2022YFF0902703), 
the National Natural Science Foundation of China (Grant No.62472121), 
the National Natural Science Foundation of China (Grant No.62306087), 
and the Special Funding Program of Shandong Taishan Scholars Project.

\newpage
\newpage
\bibliography{custom}

\newpage
\section*{Appendix}
\label{sec:appendix}
\begin{table}[!h]
  \centering 
    \resizebox{\linewidth}{!}{
    \begin{tabular}{llccc}
      \toprule
      \textbf{Provider} & \textbf{LLM} & \textbf{Input Cost}  & \textbf{Output Cost} \\
      \midrule
      \multirow{3}[1]{*}{{OpenAI}} & gpt-4 & \$30.0 & \$60.0  \\
      & gpt-4-turbo & \$10.0 & \$30.0\\
      & gpt-3.5-turbo-1106 & \$1.00 & \$2.00  \\
      \midrule
      \multirow{2}[1]{*}{{Anthropic}} & Claude-2.0 & \$11.02 & \$32.68 \\
      & Claude-instant-1.2 & \$1.63 & \$5.51  \\
      \midrule
      \multirow{3}[1]{*}{{AI21}} & Jurassic-2 Ultra & \$15.0 & \$15.0 \\
      & Jurassic-2 Mid & \$10.0 & \$10.0 \\
      & Jurassic-2 Light & \$3.00 & \$3.00 \\
      \midrule
      \multirow{7}[2]{*}{{Textsynth}} & M2M100 1.2B & \$0.15 & \$3.00 \\
      & GPT-J 6B & \$0.20 & \$5.00        \\
      & Falcon 7B & \$0.20 & \$5.00        \\
      & Mistral 7B & \$0.20 & \$2.00        \\
      & Llama2 7B & \$0.20 & \$2.00        \\
      & Flan-T5-XXL & \$0.20 & \$5.00        \\
      & Falcon 40B & \$3.30 & \$10.00       \\
      \midrule
      \multirow{2}[1]{*}{{Cohere}} & command  & \$1.00 & \$2.00 \\
      & command-light & \$0.30 & \$0.60   \\
      \midrule
      
      \multirow{8}[2]{*}{{Baidu}} & Llama-2-13B-Chat &  ￥6.00  &  ￥6.00  \\
      & Llama-2-70B-Chat &  ￥35.0  &  ￥35.0  \\
      & ERNIE-Bot 4.0 &  ￥150  &  ￥300  \\
      & ChatGLM2-6B-32K &  ￥4.00  &  ￥4.00  \\
      & Llama-2-7B-Chat &  ￥4.00  &  ￥4.00  \\
      & ERNIE-Bot &  ￥12.0  &  ￥12.0  \\
      & BLOOMZ-7B &  ￥4.00  &  ￥4.00  \\
      & ERNIE-Bot-turbo-0922 &  ￥8.00  &  ￥12.0  \\
      \bottomrule
    \end{tabular}%
    }
  \caption{Price list of different LLM services. The cost is priced per 1 million tokens. Typically, the cost of invoking LLM services consists of two components: (1) input cost (proportional to the length of
the input prompt), (2) output cost (proportional to the length of the generated sequence). Note that Baidu's LLM services are priced in Chinese Yuan (￥), while the other LLM services are priced in US Dollars (\$). The data updated to May 2024.}
  \label{price}
\end{table}%




\end{document}

%% file: taxonomy.tex
\tikzset{my-box/.style={
    rectangle,
    draw=hidden-draw,
    rounded corners,
    text opacity=1,
    minimum height=1.5em,
    minimum width=5em,
    inner sep=2pt,
    align=center,
    fill opacity=.5,
    line width=0.8pt,
}}
\tikzset{leaf/.style={my-box, minimum height=1.5em,
    fill=yellow!8, text=black, 
    align=left, font=\normalsize,
    inner xsep=2pt,
    inner ysep=4pt,
    line width=0.8pt,
}}

\begin{figure*}[t!]
    \centering
    \resizebox{0.90\textwidth}{!}{
        \begin{forest}
            forked edges,
            for tree={
                grow=east,
                reversed=true,
                anchor=base west,
                parent anchor=east,
                child anchor=west,
                base=left,
                font=\large,
                rectangle,
                draw=hidden-draw,
                rounded corners,
                align=left,
                minimum width=4em,
                edge+={darkgray, line width=1pt},
                s sep=3pt,
                inner xsep=2pt,
                inner ysep=3pt,
                line width=0.8pt,
                ver/.style={rotate=90, child anchor=north, parent anchor=south, anchor=center},
            },
            where level=1{text width=7.4em,font=\normalsize,}{},
            where level=2{text width=8.5em,font=\normalsize,}{},
            where level=3{text width=5.8em,font=\normalsize,}{},
            where level=4{text width=5em,font=\normalsize,}{},
            [
                Effective Invocation Methods, ver
                [  
                    Input \\ Abstraction (\S \ref{section:abs})
                    [
                        Sentence \\ Simplification (\S \ref{section:Sen})
                        [
                            Extractive \\ Methods 
                            [
                                TCRA-LLM~\cite{38}{, } Mondrian~\cite{8}{, } \\ Learned Token Pruning~\cite{2}{, }
                                , leaf, text width=31em
                            ]
                        ]
                        [
                            Generative \\ Methods 
                            [
                                 Commercia Models~\cite{27}{, }R0-FoMo~\cite{38}{,} \\ OverPrompt~\cite{52}
                                , leaf, text width=30em
                            ]
                        ]        
                    ]
                    [
                        Prompt \\ Optimization (\S \ref{section:Pro})
                        [
                            Prompt \\ Selection 
                            [
                                LeanContext~\cite{4}{, } \\
                                Cost-EffectiveL~\cite{37}{, } Frugal-Prompting~\cite{41}
                                , leaf, text width=31em
                            ]
                        ]
                        [
                            Prompt \\ Augmentation
                            [
                                Black-Box Tuning~\cite{11}{, }Cost Effective Testing~\cite{37}{, }\\Vision Transformer~\cite{39}{, }Factual Consistency~\cite{43}{, }\\Chain-of-Thought~\cite{44}
                                , leaf, text width=34em
                            ]
                        ]
                    ]
                ]  
                [
                    Semantic \\ Cache (\S \ref{section:Cac})
                    [
                        Traditional \\ Cache (\S \ref{section:Tra})
                        [
                            GPTCache~\cite{5}{, }Retrieval-based Dialogues~\cite{13}{, }\\Service-Caching~\cite{46}{, }Optimal-Caching~\cite{47}
                            , leaf, text width=34em
                        ]
                    ]
                    [
                        Neural \\ Cache  (\S \ref{section:Neu})
                        [
                           Cache-Distil~\cite{6}{, } VaryGen~\cite{36}{, } \\ Retrieval-based Dialogues~\cite{13}, leaf, text width=30em
                        ]
                    ]
                ]
                [
                    Solution \\ Design (\S \ref{section:int})
                       [
                          Scoring \\ Function (\S \ref{section:Eva})
                          [
                            Defined \\ Metrics[
                            Cache-Distil~\cite{6}{, }MOT~\cite{10}{, } \\ Optimal Caching~\cite{47}{, }Reward-guided~\cite{51}, leaf, text width=30em
                                ]
                            ]
                            [
                            Scorers[
                                FrugalGPT~\cite{1}{, }FORC~\cite{7}{, }\\Model-Routing~\cite{53}{, }EcoAssistant~\cite{48}{, }\\HYBRID LLM~\cite{15}{, }AutoMix~\cite{50}
                                , leaf, text width=36em
                                ]
                            ]
                    ]
                    [
                        LLM \\ Router (\S \ref{section:Rou})
                        [
                            Sequential \\ Structure 
                            [
                                FrugalGPT~\cite{1}{, }Cache-Distil~\cite{6}{, }\\MOT~\cite{10}{, }EcoAssistant~\cite{48}
                                , leaf, text width=35em
                            ]
                        ]
                        [
                            Other \\ Structure 
                            [
                                    LLM-Blender~\cite{49}{, }BRANCH-SOLVE-MERGE~\cite{57}{, }\\FORC~\cite{7}{, }Reward-guided~\cite{51}{, }\\AutoMix~\cite{50}{, }MCDM~\cite{55}{, }\\Service selection~\cite{56}
                                    , leaf, text width=37em
                            ]
                        ]
                    ]
                ]
                [
                    Output \\ Enhancement (\S \ref{section:Out})
                    [
                        Thought \\
                        Reasoning(\S \ref{section:Tho})
                        [
                            Prompting Survey~\cite{40}{, }Navigate Survey~\cite{60}{, }\\ Model Alignment~\cite{58}
                            , leaf, text width=34em
                        ]
                    ]
                    [
                        Ensemble \\
                        Learning  (\S \ref{section:Ens})
                        [
                           Ensemble Challenges~\cite{64}{, } Model Merging~\cite{65}{, } \\ Medical QA~\cite{63}{, } API Selection ~\cite{9} 
                   ni        , leaf, text width=40em
                        ]
                    ]
                ]
            ]
        \end{forest}
    }
    \caption{Taxonomy of effective invocation methods of LMaaS}
    \label{taxo_of_icl}
\end{figure*}